\documentstyle[12pt,amsfonts,axodraw]{article}
\begin{document}
\title{\bf On noncommutative and commutative equivalence for BFYM theory 
: Seiberg--Witten map } 
\author{ H. B. Benaoum  \\  
Institut f\"ur Physik , Theoretische Elementarteilchenphysik, \\      
Johannes Gutenberg--Universit\"at, 55099 Mainz, Germany \\
email : benaoum@thep.physik.uni-mainz.de
} 
\date{30 March 2000 }
\maketitle
~\\
\abstract{ BFYM on commutative and noncommutative ${\mathbb{R}}^4$ is 
considered and a Seiberg--Witten gauge--equivalent transformation is 
constructed for these theories. Then we write the noncommutative 
action in terms of the ordinary fields and show that it is 
equivalent to the ordinary action up to higher dimensional gauge 
invariant terms.
}
~\\
~\\
{\bf MZ--TH/00--12 } 
~\\
\newpage
~\\  
It is by now clear that our naive picture on space--time, as space with 
commuting coordinates, has to be modified at the Planck scale. Indeed a 
unification between quantum mechanics and general relativity requires to 
go beyond this picture. \\
Recently it has been found that noncommutative geometry~\cite{con} arises  
naturally in string theory~\cite{dou, sei}. 
In a certain limit, the low energy effective theories 
is intimately connected to gauge theories on noncommutative spaces. \\
The action for gauge theories on noncommutative space is obtained from 
the ordinary ( commutative ) Yang--Mills theories by replacing the usual 
product of the fields by the $\ast$--product~\cite{sei}. \\
It has also been shown in~\cite{sei}, that one can get ordinary or 
noncommutative Yang--Mills theory from the same two dimensional field 
theory just by using two different regularizations procedure. Indeed, 
Yang--Mills theory on noncommutative space results from the use of a 
point splitting regularization whereas the ordinary one follows as a 
consequence of the use of Pauli--Villars regularization scheme. \\
Consequently, they argued that there should exist a transformation 
from ordinary to noncommutative Yang--Mills which maps the noncommutative 
field $\hat{A}$ with gauge transformation $\hat{\lambda}$ to the ordinary 
gauge field $A$ with gauge transformation $\lambda$. More details about 
this equivalence between the two descriptions can be found in~\cite{sei} 
( see also~\cite{ish} --\cite{wes}~). \\ 
Perturbative aspects of noncommutative theories on noncommutative 
${\mathbb{R}}^d$ have been recently adressed in~\cite{mar}--\cite{min}. 
In particular, it has been found that noncommutativity leads to a strange 
mixing between IR and UV effects~\cite{min} which has no analog in the 
usual quantum field theory. \\ 
In~\cite{ben}, we have introduced the $U(1)$ BF--Yang Mills ( BFYM ) 
on noncommutative ${\mathbb{R}}^4$ and in this formulation the 
$U(1)$ Yang--Mills theory on noncommutative ${\mathbb{R}}^4$ is seen as 
a deformation of the pure BF theory. We recall that BFYM on commutative 
${\mathbb{R}}^4$ has been introduced and studied extensively 
in~\cite{fuc}. \\
We have also performed a one--loop calculation of the $U(1)$ BFYM on 
noncommutative ${\mathbb{R}}^4$ and concluded that it is asymtotically 
free and its UV--behaviour in the computation of the $\beta$--function is 
like the usual $SU( N )$ commutative BFYM and Yang--Mills theories. \\
Here we will go a step further and study the equivalence between 
the noncommutative and commutative for BFYM. We consider first rank $N$ 
noncommutative and ordinary BFYM and construct the gauge--equivalent 
Seiberg--Witten transformation. This means that ordinary and 
noncommutative BFYM theories can be induced by the same field theory 
regularized in two different ways. \\
In particular for rank 1, this mapping is more simple and for this 
case, we write the BFYM noncommutative Lagrangian in terms of the 
ordinary fields. \\ 
~\\
Noncommutative ${\mathbb{R}}^4$ is described by an algebra generated by 
the coordinates $x^i~( i = 1, \cdots 4 )$ obeying the commutation 
relations :
\begin{eqnarray}
x^i \ast x^j - x^j \ast x^i & = & i~\theta^{i j} 
\end{eqnarray}
where $\theta^{i j} = - \theta^{j i}$ is a real and constant antisymmetric 
matrix. In the dual language, the algebra of functions on ${\mathbb{R}}^4$ 
vanishing at infinity is equipped with a deformed multiplication known as 
a Moyal $\ast$--product,  
\begin{eqnarray}
( f \ast g ) ( x ) & = & e^{\frac{i}{2}~\theta^{i j}
\frac{\partial}{\partial \xi^{i}} \frac{\partial}{\partial \zeta^{j}}}~
f( x + \xi ) g( x + \zeta) |_{\xi = \zeta = 0} 
\end{eqnarray} 
which is noncommutative, associative and obeys the Leibniz rule. It   
reduces for the first order in 
$\theta$ to, 
\begin{eqnarray}
( f \ast g ) ( x ) & = & f g + \frac{i}{2}~\theta^{i j}~\partial_i f 
\partial_j g + O\left( \theta^2 \right) 
\end{eqnarray}  
~\\
We consider the BFYM on noncommutative ${\mathbb{R}}^4$ introduced in ~
\cite{ben} by replacing the usual product of fields by the $\ast$--product, 
\begin{eqnarray}
\hat{S} & = & \int d^4 x~Tr \left( \frac{i}{2}~ \epsilon^{i j k l}
\hat{B}_{i j} \ast \hat{F}_{k l} + g^2 \hat{B}_{i j} \ast \hat{B}^{i j}
\right) ( x ).
\end{eqnarray} 
where $\hat{F}_{i j} = \partial_i \hat{A}_j - \partial_j \hat{A}_i + 
\hat{A}_i \ast \hat{A}_j - \hat{A}_j \ast \hat{A}_i$ is the field strength of 
the antihermitian gauge field $\hat{A}_i$ and $\hat{B}_{i j}$ is an 
antisymmetric tensor. We take here the gauge fields $\hat{A}_i$ and 
$\hat{B}_{i j}$ to be arbitrary rank $N$, so that all the fields and gauge 
parameters are $N \times N$ matrices and $Tr$ is the ordinary trace of 
$N \times N$ matrices. \\ 
This BFYM on noncommutative ${\mathbb{R}}^4$ is invariant under the gauge 
transformation, 
\begin{eqnarray}
\hat{\delta}_{\hat{\lambda}} \hat{A}_i & = & \partial_i \hat{\lambda} + 
\hat{\lambda} \ast \hat{A}_i - \hat{A}_i \ast \hat{\lambda} \nonumber \\
\hat{\delta}_{\hat{\lambda}} \hat{B}_{i j} & = & \hat{B}_{i j} \ast 
\hat{\lambda} - \hat{\lambda} \ast \hat{B}_{i j} \nonumber \\
\hat{\delta}_{\hat{\lambda}} \hat{F}_{i j} & = & \hat{F}_{i j} \ast 
\hat{\lambda} - \hat{\lambda} \ast \hat{F}_{i j} 
\end{eqnarray} 
Using the first order expansion in $\theta$ of the Moyal $\ast$--product 
( 3), the above formulas for noncommutative gauge field $\hat{A}_i$ and 
$\hat{B}_{i j}$ read,   
\begin{eqnarray} 
\hat{\delta}_{\hat{\lambda}} \hat{A}_i & = & \partial_i \hat{\lambda} + 
[ \hat{\lambda}, \hat{A}_i ] +  
\frac{i}{2}~\theta^{k l}~\{ \partial_k \hat{\lambda}, 
\partial_l \hat{A}_i \} + O \left( \theta^2 \right) \nonumber \\
\hat{\delta}_{\hat{\lambda}} \hat{B}_{i j} & = & [ \hat{B}_{i j}, \hat{\lambda} ] 
+ \frac{i}{2}~\theta^{k l}~\{ \hat{B}_{i j}, \hat{\lambda} \} + 
O \left( \theta^2 \right) 
\end{eqnarray}  
where $[~,~]$ and $\{~,~\}$ are the usual commutator and anticommutator 
for noncommutative fields and gauge parameter $\hat{\lambda}$ respectively. \\ 
Similarly, the BFYM on commutative ${\mathbb{R}}^4$ is, 
\begin{eqnarray} 
S & = & \int d^4 x~Tr \left( \frac{i}{2}~ \epsilon^{i j k l}
B_{i j} F_{k l} + g^2 B_{i j} B^{i j}
\right) ( x ).
\end{eqnarray}
Its infinitesimal gauge transformations for the different fields are of the 
form, 
\begin{eqnarray} 
\delta_{\lambda} A_i & = & \partial_i \lambda + [ \lambda, A_i ] \nonumber \\
\delta_{\lambda} B_{i j} & = & [ B_{i j}, \lambda ] \nonumber \\
\delta_{\lambda} F_{i j} & = & [ F_{i j}, \lambda ]
\end{eqnarray} 
~\\
Following Seiberg and Witten~\cite{sei}, we consider the BFYM and look for 
a gauge--equivalent mapping between the noncommutative fields $\hat{A}_i$ and 
$\hat{B}_{i j}$ with gauge parameter $\hat{\lambda}$ and the ordinary    
fields $A_i$ and $B_{i j}$ with gauge 
parameter $\lambda$ such that the following diagrams are commutative : 
\begin{center}
\begin{picture}(400,50)
\thicklines 
\LongArrow(40,20)(120,20)
\LongArrow(30,-30)(30,10)
\LongArrow(40,-40)(120,-40)
\LongArrow(140,-30)(140,10) 
\LongArrowArcn(80,-10)(15,-30, 60)
\Text(20,-5)[]{$\hat{\delta}_{\hat{\lambda}}$}
\Text(150,-5)[]{$\delta_{\lambda}$} 
\Text(20,20)[]{$\hat{\delta}_{\lambda} \hat{A}_i$}
\Text(30,-40)[]{$\hat{A}_i$}
\Text(140,-40)[]{$A_i$}
\Text(140,20)[]{$\delta_{\lambda} A_i$} 
\LongArrow(220,20)(300,20)
\LongArrow(210,-30)(210,10)
\LongArrow(220,-40)(300,-40)
\LongArrow(320,-30)(320,10) 
\LongArrowArcn(270,-10)(15,-30,60)
\Text(200,-5)[]{$\hat{\delta}_{\hat{\lambda}}$}
\Text(330,-5)[]{$\delta_{\lambda}$}
\Text(200,20)[]{$\hat{\delta}_{\hat{\lambda}} \hat{B}_{i j}$}
\Text(210,-40)[]{$\hat{B}_{i j}$}
\Text(320,-40)[]{$B_{i j}$}                                                  
\Text(320,20)[]{$\delta_{\lambda} B_{i j}$} 
\end{picture} \\
\end{center}  
~\\
~\\
~\\
This means that the two following relations hold : 
\begin{eqnarray} 
\hat{A}_i ( A ) + \hat{\delta}_{\hat{\lambda}} \hat{A}_i ( A ) & = & 
\hat{A}_i ( A + \delta_{\lambda} A ) \nonumber \\
\hat{B}_{i j} ( B ) + \hat{\delta}_{\hat{\lambda}} \hat{B}_{i j} ( B ) & = & 
\hat{B}_{i j} ( B + \delta_{\lambda} B ) 
\end{eqnarray}
By writing $\hat{A} = A + A' ( A ) + O \left( \theta^2 \right), 
\hat{\lambda} ( \lambda , A ) = \lambda + \lambda' ( \lambda, A) + 
O \left( \theta^2 \right), 
\hat{B} = B + B' ( B, A ) + O \left( \theta^2 \right)$ and 
expanding (9 ) to the first order in $\theta$, we get for gauge field,  
\begin{eqnarray} 
A'_i ( A + \delta A) - A'_i ( A ) - \partial_i \lambda' - 
[ \lambda', A_i ] - [ \lambda, A'_i ] & = & 
- \frac{i}{2}~\theta^{k l} \{ \partial_k \lambda, \partial_l A_i \} + 
O \left( \theta^2 \right) 
\end{eqnarray} 
and for antisymmetric field, 
\begin{eqnarray}
B'_{i j} ( B + \delta_{\lambda} B ) - B_{i j} ( B ) - [ B_{i j}, \lambda' ] - 
[ B'_{i j}, \lambda ] & = & 
- \frac{i}{2}~\theta^{k l} \{ \partial_k B_{i j}, \partial_l \lambda \} + 
O \left( \theta^2 \right). 
\end{eqnarray} 
By solving (10) and (11), the following solutions for the noncommutative 
gauge field $\hat{A}$, gauge parameter $\hat{\lambda}$ and antisymmetric field 
$\hat{B}$ are obtained : 
\begin{eqnarray}
\hat{A}_i ( A ) & = & A_i + A'_i ( A ) = A_i - \frac{i}{4}~\theta^{k l}~
\{ A_k , \partial_l A_i + F_{l i} \} + O \left( \theta^2 \right) \nonumber \\
\hat{\lambda} ( \lambda, A ) & = & \lambda + \lambda' ( \lambda, A ) = 
\lambda + \frac{i}{4}~\theta^{k l} \{ \partial_k \lambda, A_l \} + 
O \left( \theta^2 \right) \nonumber \\
\hat{B}_{i j} ( B, A ) & = & B_{i j} + B'_{i j} ( B, A ) \nonumber \\ 
& = & B_{i j} + \frac{i}{4}~\theta^{k l}~\left(~2~\{ B_{i k}, B_{j l} \}  
- \{ A_k, D_l B_{i j} + \partial_l B_{i j} \} \right) + 
O \left( \theta^2 \right) .
\end{eqnarray} 
where $D_l B_{i j} = \partial_l B_{i j} + [ A_l, B_{i j} ]$. \\
From the gauge field $\hat{A}_i$, the noncommutative   
field strength $\hat{F}_{i j}$ is 
given as : 
\begin{eqnarray}
\hat{F}_{i j} & = & F_{i j} + \frac{i}{4}~\theta^{k l}~
\left(~2~\{ F_{i k}, F_{j l} \} -
\{ A_k, D_l F_{i j} + \partial_l F_{i j} \} \right) +
O \left( \theta^2 \right) . 
\end{eqnarray} 
This means that knowing the commutative fields $A_i, B_{i j}$ and 
$F_{i j}$ with gauge parameter $\lambda$, the noncommutative one are 
expressed by (12) and (13) up to first order in $\theta$. \\
Now it is easy to extract the differential equation for noncommutative 
fields $\hat{A}, \hat{B}$ and $\hat{F}$ with gauge parameter 
$\hat{\lambda}$ corresponding to two infinitesimally close values 
of the deformation parameter $\theta$ and $\theta + \delta \theta$. \\
They are given as follows : 
\begin{eqnarray}
\delta \hat{A}_i & = & - \frac{i}{4}~\delta \theta^{k l}~[~ 
\hat{A}_k \ast ( \partial_l \hat{A}_i + \hat{F}_{l i} ) + 
( \partial_l \hat{A}_i + \hat{F}_{l i} ) \ast \hat{A}_k ~] \nonumber \\
\delta \hat{\lambda } & = & \frac{i}{4}~\delta \theta^{k l}~( 
\partial_k \hat{\lambda} \ast \hat{A}_l + \hat{A}_l \ast 
\partial_k \hat{\lambda} ) \nonumber \\
\delta \hat{B}_{i j} & = & \frac{i}{4}~\delta \theta^{k l}~[~ 
2 \hat{B}_{i k} \ast \hat{B}_{j l} + 2 \hat{B}_{j l} \ast \hat{B}_{i k} - 
\hat{A}_k \ast ( \hat{D}_l \hat{B}_{i j} + \partial_l \hat{B}_{i j} ) 
\nonumber \\ 
& & - ( \hat{D}_l \hat{B}_{i j} + \partial_l \hat{B}_{i j} ) \ast \hat{A}_k~] 
\nonumber \\
\delta \hat{F}_{i j} & = & \frac{i}{4}~\delta \theta^{k l}~[~
2 \hat{F}_{i k} \ast \hat{F}_{j l} + 2 \hat{F}_{j l} \ast \hat{F}_{i k} -
\hat{A}_k \ast ( \hat{D}_l \hat{F}_{i j} + \partial_l \hat{F}_{i j} )
\nonumber \\
& & - ( \hat{D}_l \hat{F}_{i j} + \partial_l \hat{F}_{i j} )  \ast \hat{A}_k~] 
\end{eqnarray} 
where $\hat{D}_l \hat{B}_{i j} = \partial_l \hat{B}_{i j} + 
\hat{A}_l \ast \hat{B}_{i j} - \hat{B}_{i j} \ast \hat{A}_l$. \\
The above equations should in principle give the noncommutative fields 
and gauge parameter as powers series in $\theta$. \\
~\\
For rank 1, i.e. $U( 1)$, the Seiberg--Witten map (12) for different fields 
reduces to :    
\begin{eqnarray}
\hat{A}_i & = & A_i - \frac{i}{2} \theta^{k l} A_k ( \partial_l A_i + 
F_{l i} ) + O \left( \theta^2 \right) \nonumber \\
\hat{\lambda} & = & \lambda + \frac{i}{2} \theta^{k l} \partial_k \lambda 
A_l + O \left( \theta^2 \right) \nonumber \\ 
\hat{B}_{i j} & = & B_{i j} + i \theta^{k l} ( B_{i k} B_{j l} - A_k 
\partial_l B_{i j} ) + O \left( \theta^2 \right) \nonumber \\ 
\hat{F}_{i j} & = & F_{i j} + i \theta^{k l} ( F_{i k} F_{j l} - A_k 
\partial_l F_{i j} ) + O \left( \theta^2 \right)  
\end{eqnarray} 
~\\ 
The noncommutative action $\hat{S}$ in terms of the fields 
$\hat{A}, \hat{B}$ and $\hat{F}$ and the commutative one $S$ in 
terms of the fields $A, B$ and $F$ for BFYM are related as : 
\begin{eqnarray}
\hat{S} & = & \int d^4 x~ \hat{ \cal L} (\hat{B}, \hat{F} ) \nonumber \\ 
& = & \int d^4 x~ \{~\frac{i}{2}~\epsilon^{i j k l} 
B_{i j} F_{k l} + g^2 B_{i j} 
B^{i j} - \frac{1}{2}~\theta^{n m}~\epsilon^{i j k l} F_{k l} B_{i n} 
B_{j m} \nonumber \\ 
& & + \frac{1}{2}~\theta^{n m}~\epsilon^{i j k l} B_{i j} F_{n m} 
F_{k l} - \frac{1}{2}~\theta^{n m}~\epsilon^{i j k l} B_{i j} F_{k n} 
F_{l m} + 2 i g^2 \theta^{n m} B^{i j} B_{i n} B_{j m} \nonumber \\
& & - i g^2 \theta^{n m} B^{i j} F_{n m} B_{i j} 
+ total~derivative~+ O\left( \theta^2 \right)~ \} \nonumber \\ 
& = & \int d^4 x~ \frac{1}{1 + i F \theta}~\{~\frac{i}{2}~\epsilon^{i j k l} 
\left( \frac{1}{1 + i B \theta}~B \right)_{i j}~\left( 
\frac{1}{1 + i F \theta}~F \right)_{k l} \nonumber \\ 
& & + g^2~\left( \frac{1}{1 + i B \theta}~B \right)_{i j}~
\left( \frac{1}{1 + i B \theta}~B \right)^{i j}  + 
total~derivative~+ O\left( \theta^2 \right)~ \}  
\end{eqnarray} 
The $U(1)$ BFYM theory on noncommutative ${\mathbb{R}}^4$ 
with noncommutative fields $\hat{B}, \hat{F}$ written in terms 
of the ordinary fields as in (16) is equivalent to $U(1)$ BFYM 
on commutative ${\mathbb{R}}^4$ with antisymmetric field 
$\frac{1}{1 + i B \theta}~B$ and strength field $\frac{1}{1 + i F \theta}
~F$ where the factor $\frac{1}{1 + i F \theta}$ results from the 
partial derivation on the fields $B$ and $F$ in (15). \\  
The solutions, 
\begin{eqnarray}
\hat{F}~=~\frac{1}{1 + i F \theta}~F~,~~~
\hat{B} = \frac{1}{1 + i B \theta}~B.
\end{eqnarray} 
follows also from the resolution of the differential equation (14) for 
constant fields $\hat{B}, \hat{F}$. \\
For this case, (16) can be written as : 
\begin{eqnarray}
\hat{S} & = & \int d^4 x~\hat{ \cal L} ( \hat{B}, \hat{F} ) \nonumber \\
& = & \int d^4 x~\{~{\cal L} ( \frac{1}{1 + i B \theta}~B, 
\frac{1}{1 + i F \theta}~F ) + O \left( \theta^2 \right)~\}.
\end{eqnarray} 
~\\
We close this paper by saying that the existence of a Seibeg--Witten 
gauge--equivalent map, relating noncommutative fields to 
commutative one for BFYM theory, means that noncommutative or ordinary 
BFYM may be seen as an effective theory which arises from the same field 
theory regularized with two different schemes. 
\section*{Acknowledgment} 
I would like to thank the DAAD for its financial support and Y. Okawa for 
his comments and remarks. 

\end{document}